# Demonstration of a Single-Laser-Diode-Pumped Ti:Sapphire Astrocomb on the Southern African Large Telescope

Ewan Allan[1], Yuk Shan Cheng[1], Kamalesh Dadi[1], Pablo Castro-Marin[1], Jake M. Charsley[1], William Newman[1], Abdullah Alabbadi[2,3,4], Hanna Ostapenko[1], Richard A. McCracken[1], Pascal Del'Haye[2,4], Thomas Willemsen[5], Tobias Gross[5], Daniel L. Holdsworth[6,7,8], Malcolm C. Scarrott[6,7], Lisa A. Crause[6,7], and Derryck T. Reid[1]*

*D.T.Reid@hw.ac.uk

1. Institute of Photonics and Quantum Sciences, Heriot-Watt University, Edinburgh, UK.
2. Max Planck Institute for the Science of Light, Erlangen, Germany.
3. Department of Physics, Alexandria University, Alexandria, Egypt.
4. Department of Physics, Friedrich-Alexander-Universität Erlangen-Nürnberg, Germany.
5. LASEROPTIK GmbH, Johannes-Ebert-Str. 1, 30826 Garbsen, Germany.
6. South African Astronomical Observatory, Cape Town, South Africa.
7. Southern African Large Telescope, South Africa.
8. School of Physics, Engineering and Technology, University of York, Heslington, York, UK.

**Abstract:** Astrocombs—broadband lasers comprising thousands of narrow, uniformly spaced and atomically-referenced spectral lines—offer gold-standard wavelength calibration for ground-based optical telescopes, with the potential for the cm $s^{-1}$ precision needed for radial-velocity follow-ups of exoplanet candidates from photometric space missions like Kepler, TESS and PLATO. Current astrocombs, particularly those operating in the visible spectrum, are complex and expensive, putting them beyond the reach of many observatories. Here, we present the first on-instrument demonstration of a new and simple astrocomb concept, providing calibration light across nearly the entire 550 nm to 890 nm red channel of the Southern African Large Telescope's High Resolution Spectrograph (SALT-HRS). An octave-spanning supercontinuum generated in a silicon nitride waveguide by a GPS-referenced diode-pumped Ti:sapphire laser is filtered to 21 GHz before fibre-delivery to the spectrograph. Using the astrocomb, we obtain the wavelength solution for one order of SALT-HRS and the wavelength-dependent line-spread function of the instrument. With potential for further extension into the blue spectral region, this uniquely simple architecture brings visible-to-infrared astrocomb technology within the reach of a wider range of astronomical observatories.

## Introduction

Ground-based radial-velocity (RV) measurements are essential for planetary mass confirmation of exoplanet candidates from transit studies like Kepler [1], TESS [2] and PLATO [3]. The long orbital periods and tiny RV shifts from Earth-like planets demand high precision, accuracy and traceability in astrophysical spectrographs, motivating the adoption of laser frequency combs (LFCs), a calibration technology providing thousands of ultra-narrow, atomically-referenced optical frequencies, uniformly spaced at multi-GHz intervals. Visible-range astrocombs are typically expensive (>$1M), complex and difficult to maintain remotely, limiting deployment to perhaps 20 observatories worldwide. Reliable deployable yet simple astrocombs are therefore critically needed for the benefits of this gold-standard calibration technology to have the widest impact on the exoplanet observation community, including ambitions to adopt LFCs as calibrators in the portfolio of candidate extreme-precision RV telescopes recommended by the NASA / NSF EPRV Working Group [4].

In the nearly two decades since the first astrocomb demonstration [5], a variety of architectures has been reported, including those based on modelocked fibre lasers [6-8] and Ti:sapphire lasers [9-13], and more recently on electro-optical combs [14-17] and micro-resonator combs [18-20]. In particular, Ti:sapphire lasers are attractive platforms for astrocombs that use Fabry-Pérot filtering because of their 1 GHz [9-12] or higher [13] native comb spacings and the intrinsically low phase noise and narrow comb linewidths available from solid-state laser oscillators [21]. The near-visible 800 nm wavelength of Ti:sapphire lasers has also made them a popular choice for astrocombs targeting blue / green wavelengths [11,13,22-24]. A dramatic simplification is provided by directly diode-pumped femtosecond Ti:sapphire lasers [25,26], which can be configured as fully stabilized LFCs [27], offering comparable performance to those pumped by considerably more complex frequency-doubled diode-pumped solid-state (DPSS) lasers, including recent extensions to GHz-rate operation and multi-100 mW average output powers [28-31,32]. Here, we present the first example of the deployment in an observatory setting of a diode-pumped Ti:sapphire astrocomb, providing 21 GHz comb modes with nearly complete visible-to-infrared coverage of the 550 nm to 890 nm red-channel of the High-Resolution Spectrograph (HRS) on the Southern African Large

Telescope (SALT). By eliminating complex and high-cost pump lasers and amplifier chains, this demonstration represents a radical simplification of a deployable visible-band astrocomb.

## Results

### *Astrocomb configuration*

The astrocomb is based on a 1 GHz Kerr-lens-modelocked diode-pumped Ti:sapphire laser (Fig. 1, dashed black box), similar in design to that described in [28] and with the addition of coarse and fine remote adjustment of the cavity length to allow repetition-rate stabilisation. A few mW of the laser output is sampled using a beam-splitter (BS) to obtain its pulse repetition frequency, which is used to stabilise the primary comb-mode spacing ($f_{\mathrm{rep}}$) to a 1.003 GHz reference oscillator, disciplined to the onboard caesium clocks of the GPS satellite constellation with a stability approaching $10^{-12}$ [33]. A servo loop (Fig. 1a-g) phase-locks the laser repetition frequency to the GPS reference, stabilizing the comb-mode spacing to < 400 μHz over a few seconds, and with slow correction to the cavity length this lock can be maintained indefinitely. When the comb's mode spacing is controlled in this way, the only remaining free variable in the mode positions is their offset frequency, which exhibits a thermally-driven 9.4 MHz rms variation over one hour, and can be directly measured by a heterodyne beat measurement with a single-frequency continuous-wave (cw) laser. The rms variation over typical spectrograph exposure times of up to 300 seconds is < 7.9 MHz, corresponding to an RV uncertainty at 780 nm of 6.2 m s$^{-1}$, which is comparable to the original baseline design specification for the SALT-HRS spectrograph of ≈ 4 m s$^{-1}$. Therefore, even without active comb-offset control, the comb-offset stability is sufficient, however this would be improved to Hz-level stability by using self-referencing [34] to lock the comb carrier-envelope offset frequency ($f_{\mathrm{CEO}}$) to a constant value.

An octave-spanning supercontinuum from 540 nm to 1080 nm and with around 4 mW of average output power is produced by pumping a silicon nitride ($Si_3N_4$) waveguide, fabricated with a 15° facet angle to avoid feedback [35]. The supercontinuum is coupled into a broadband polarisation-maintaining fibre (Thorlabs PM630) which serves as a spatial filter to homogenize the beam leaving the waveguide and permits deterministic mode-matching into the Fabry-Pérot filter cavity used to create the astrocomb. The plano-concave filter cavity has a free-spectral range of $21f_{\mathrm{rep}}$ (21,063,096,600 Hz), corresponding to a physical mirror separation of approximately 7.1 mm. The 98% reflectivity mirrors have complementary group-delay dispersion characteristics that provide a nearly flat group-delay from 550 nm to 900 nm with an rms variation per reflection of only 0.08 fs, providing >20 dB suppression of the primary LFC comb modes across this range. The output from the Fabry-Pérot cavity is first coupled into a 10-metre P3-630PM-FC-10 fibre then delivered to the SALT-HRS spectrograph launch bench through a 10-metre large-mode-area, polarization-maintaining photonic crystal fibre (NKT LMA-PM-5). This fibre is chosen because it provides single-mode operation with low loss and near-constant mode-field diameter across the entire spectrograph bandwidth. A dither-locking scheme maintains the Fabry-Pérot transmission on a maximum by using an avalanche photodiode (APD) to sample the power transmitted near 800 nm to produce an error signal in an FPGA-based servo loop (Fig. 1h-n).

***Performance on the SALT HRS spectrograph***

SALT-HRS is a two-channel cross-dispersed echelle spectrograph that uses a single R4 echelle grating with 41.6 grooves / mm and volume phase holographic gratings as cross-dispersers [36,37]. In the high-stability mode used here, the comb is fibre-fed to the spectrograph by two 350 μm core fibres, whose apertures are each image-sliced to match the spectrograph input slits, accounting for the apparent comb-line structure visible in the insets of Fig. 2. The effective resolving power ($\lambda/\Delta\lambda$) in the high-stability mode is ≈69,200 [36], equivalent to a frequency resolution of 4.9 GHz to 7.9 GHz from 890 nm to 550 nm respectively. Comb modes with typical linewidths of 100 kHz are unresolved by the spectrograph and their offset drift (<10 MHz rms) is undetectable, even in long exposures.

The main image of Figure 2 shows a complete frame recorded on the red channel of SALT-HRS with a ten-second exposure time in high-stability mode. The false colour rendering is provided as a visual indicator of wavelength, varying from green at 550 nm to red at 890 nm, and is not intended to be colorimetrically accurate. The intensity of the image is percentile scaled between 50% (lowest intensity) and 98% (highest intensity) to account for the exact distribution of the background intensity and to suppress extreme outlier pixels. The dashed white lines represent the free-spectral range of the spectrograph, which varies from 6.5 nm at 550 nm to 16.6 nm at 880 nm. Each order is indicated by the order number and blaze wavelength. Astrocomb light was coupled predominantly into one of the two 350-μm-core fibres, but some leakage into the second fibre channel is apparent as weaker image visible immediately above many orders.

Insets 2–6 in Fig. 2 show the detail of individual regions from across the frame. The apparent saturation and widening of some comb lines is not real, but is caused by the percentile scaling applied to optimally render the wide dynamic range of the data. An example line-cut through ten comb lines present in order number 64 (indicated by the white arrow) is shown below inset 6. Each comb line in the spectrograph data (symbols) is fitted with a Gaussian function (red) of adjustable centre, amplitude, baseline and width. The apparent absence of comb lines in the 605 nm to 665 nm portion of the frame is caused by the low spectral intensity of the $Si_3N_4$ supercontinuum in this region, however at 300 seconds exposure time comb lines become easily visible, as illustrated in Inset 1.

As a demonstration of the calibration potential of the astrocomb, Fig. 3 shows the result of a complete fit to 408 comb lines on order number 64 of the SALT-HRS frame shown in Fig. 2, where a fifth-order Legendre polynomial fit to the comb lines determines the wavelength solution for the order (Fig. 3a, red line). The precision of the wavelength solution is inferred from the fitting residuals in Fig. 3b, which are expressed in both wavelength ($\sigma_{\lambda}$ = 0.078 pm) and radial velocity ($\sigma_{\mathrm{RV}}$ = 31.9 m s$^{-1}$). Each Gaussian fit to a comb line serves as a reporter of the wavelength-dependent line-spread function of SALT-HRS, allowing the resolving power to be obtained across the order. In Fig. 3c we show the results of this analysis, which indicates a median resolving power of 66,320 that is consistent with published design values [36,37].

**Discussion**

As currently demonstrated, the comb-offset frequency shows sufficient free-running stability to enable several-m s$^{-1}$ precision measurements, with absolute referencing to a wavemeter-referenced single-frequency cw laser. Higher precision could be obtained by offset-locking the comb to the cw laser, for example by using a feed-forward technique [38], and by atomically referencing the cw laser to a Rb transition near 780 nm. Self-referencing would further improve the stability of the comb offset frequency, and could be implemented with the pulse energies available by using a suitable photonic crystal fibre [39] to provide an f-to-2f beat frequency for directly controlling $f_{\mathrm{CEO}}$ through pump-current modulation [27]. In separate work, we have obtained efficient spectral broadening from supercontinuum generation in photonic crystal fibre, opening the possibility of extension into the blue spectral region (astronomical B-band) through sum-frequency and second-harmonic generation in nonlinear waveguides [24]. The significance of this first on-instrument deployment of an astrocomb based on a single-diode-pumped laser is to demonstrate the potential of such technology to simplify astrocomb systems and so to bring B- to Z-band astrocomb-assisted calibration within the reach of a wider range of astronomical observatories.

## Methods

### *Diode-pumped Ti:sapphire laser*

The Ti:sapphire laser was adapted from a previously reported design [28,29] and pumped by a Nichia NDG7D75 broad-stripe laser diode. The oscillator produced sub-80 fs pulses; under the experimentally used waveguide conditions, noise-seeded GNLSE simulations show that this pulse-duration range supports efficient soliton-mediated dispersive-wave generation while preserving high first-order coherence across the spectral bands used for astrocomb operation (see Supplementary Figs. S1 and S2). A total output power of 153 mW was produced, of which 6.4 mW was used for repetition-rate locking and 144 mW was available for supercontinuum generation. The intracavity plane mirror was coated on a lightweight substrate, which was bonded onto a piezoelectric transducer for high-speed cavity-length modulation. The mirror was attached to a translation stage containing a Picomotor™ linear actuator with sub-30 nm adjustment and a 12.7 mm travel range, allowing the pulse repetition frequency to be adjusted in coarse steps of ≈200 Hz.

### *Laser pulse repetition-rate stabilisation*

The laser pulse repetition frequency was detected on a fast silicon PIN photodiode (Fig. 1, PIN) whose output was amplified and bandpass filtered around 1 GHz to produce a sine wave that supplied one input of a double-balanced mixer. The other input of the mixer was a 1,003,004,600 Hz reference frequency from a high-performance GPS-disciplined oscillator (Leo Bodnar LBE-1421) with a stability of $10^{-12}$ at 1000 seconds. Low-pass filtering of the mixer output to frequencies below 300 kHz provided an error signal that served as the input to a proportional-integral amplifier that was used to directly actuate a piezo-electric transducer on which an intracavity mirror was mounted.

The in-loop error signal was digitised at 5 MHz for 13 ms and at 20 kHz for 3.3 s to provide overlapping phase-noise power spectral density (PSD) measurements from which the frequency instability of the pulse repetition rate could be estimated. Figure 4 presents the reconstructed phase noise PSD (left axis) and shows that the phase noise integrated from 1 MHz to 300 mHz (right axis) amounts to 8 mrad over an observation time of 3.3 seconds. This phase error implies a comb-mode spacing frequency instability of 0.389 mHz, equivalent to an optical frequency uncertainty at 650 nm of 180 Hz. Importantly, this in-loop measurement is insensitive to variations in the reference frequency, whose specified stability is defined by the accuracy of GPS satellite onboard Caesium references and therefore implies a comb-spacing-limited optical uncertainty of ≈460 Hz.

***Measurement of comb offset stability***

With the comb repetition-frequency locked, a heterodyne beat in the range $0 < f_{\mathrm{beat}} < f_{\mathrm{rep}}/2$, was recorded between the comb and a single-frequency diode laser (Toptica DL pro 780), whose wavelength was simultaneously recorded on a High Finesse WS7-60 wavemeter with an absolute accuracy of 60 MHz and a wavelength deviation sensitivity of 2 MHz. The frequency of the diode laser, $f_{\mathrm{cw}}$, was adjusted to be 300 MHz–400 MHz above the frequency of the nearest comb mode, $f_n$, confirmed by observing that increases in $f_{\mathrm{cw}}$ also increased $f_{\mathrm{beat}}$. The frequency of the nearest comb line is given by $f_n = f_{\mathrm{cw}} - f_{\mathrm{beat}}$, with an index of $n = \lfloor f_n/f_{\mathrm{rep}} \rfloor$, giving the comb offset frequency from $f_{\mathrm{CEO}} = f_n - nf_{\mathrm{rep}}$. This approach can be applied to provide the absolute offset frequency of the astrocomb and avoids the ambiguity between $f_{\mathrm{CEO}}$ and $f_{\mathrm{rep}} - f_{\mathrm{CEO}}$ present when $f_{\mathrm{CEO}}$ is determined from an f-to-2f self-referenced beat note [34].

The beat note and wavemeter data were recorded for one hour and changes in the comb-offset frequency are presented in Fig. 5, showing a 9.4 MHz rms variation, with a characteristic 20-minute modulation caused by the lab air-conditioning unit. Over timescales of 10 s, 60 s and 300 s that are characteristic of spectrograph exposures, the rms variations are 2.1 MHz, 3.7 MHz and 7.9 MHz respectively, with corresponding RV uncertainties at 780 nm of 1.7 m $s^{-1}$, 2.9 m $s^{-1}$ and 6.2 m $s^{-1}$.

***Fabry-Pérot filter cavity stabilisation***

The plano-concave Fabry-Pérot filter cavity comprised two 12.7 mm diameter mirrors, one plane and the other with a 38 mm radius of curvature, bonded onto ring PZTs, which in turn were bonded to brass mounts (Fig. 6a). Both mirrors could be independently aligned and their separation could be adjusted using a Picomotor™-driven translation stage under one of the mounts. Fibre-coupled light was mode-matched into the Fabry-Pérot cavity using a 150-mm focal length lens to form a 60 μm radius waist (Fig. 6b) and was output coupled by a 125-mm focal length lens. With the Fabry-Pérot first aligned with a fibre-coupled 780 nm cw laser and the finesse optimised to achieve close to the theoretical value, the length resonant with an integer multiple of $f_{\mathrm{rep}}$ was found by scanning the mirror separation over several mm and observing the characteristic Vernier transmission fringes, whose pattern can be accurately predicted by theory. For an astrocomb of mode spacing $mf_{\mathrm{rep}}$, this process allowed the principal transmission peak to be found, corresponding to a mirror separation of $v_g/2mf_{\mathrm{rep}}$.

The Fabry-Pérot cavity was stabilised onto the chosen transmission peak using dither locking. A 10 kHz low-voltage signal (Fig. 1k) was applied to the PZT on the concave mirror and a phase-shifted replica of this signal (Fig. 1j) used with a Level-7 double-balanced mixer (Fig. 1i) to demodulate the amplified transmission signal (Fig. 1h). Low-pass filtering at 976 Hz (Fig. 1m) created an error signal for a Red Pitaya FPGA board (Fig. 1n) that, after a high-voltage amplifier (Fig. 1l), actuated the PZT on the plane Fabry-Pérot mirror. Initial alignment of the Fabry-Pérot was performed using a 10 Hz ramp signal that was turned off for locking. Figure 6c and 6d show, respectively, the error signal and the transmission signal with and without stabilisation.

***Silicon nitride waveguide design and performance***

The silicon nitride waveguides used in this work are described in detail elsewhere [30,35] and were fabricated using subtractive electron-beam lithography on a 400-nm-thick low-pressure chemical vapor deposition (LPCVD) $Si_3N_4$ film grown on a thermally oxidized silicon substrate (Fig. 7a). The samples were cleaved to enable edge coupling (Fig. 7b), with the lithography designed to present the waveguides at an internal tilt of 15° (Fig. 7c) to avoid feedback which would disrupt modelocking in the Ti:sapphire laser. The calculated dispersion for the quasi-TE mode of the waveguides (Fig. 7d) shows anomalous dispersion from 700 nm to 900 nm, making it suitable for supercontinuum generation when pumped near 800 nm [35]. In Fig. 7e we show example spectra from each of a group of five waveguides fabricated with a width of 600 nm, illustrating the consistency of the supercontinuum performance when pumped by 25 pJ on-chip pulse energy.

***Comb-mode fitting procedure yielding the spectrograph wavelength solution***

We identified astrocomb modes using the scipy.signal.find_peaks code in Python and fitted each detected mode with a Gaussian to precisely determine the centroid location in pixel space. Using a wavelength solution derived from a Thorium Argon hollow cathode lamp, we identified an astrocomb mode that fell in the same pixel location as a ThAr emission line close to the centre of the order to use as an absolute wavelength anchor. We calculated a theoretical comb spectrum for the order using this anchor and the GPS-referenced comb-mode spacing. Observed comb modes were matched to the theoretical ones to derive the wavelength solution for the order (Fig. 3a, diamonds) and a fifth-order Legendre polynomial fit is applied to determine the wavelength solution for the order (Fig. 3a, red line).

***Dispersion-engineered Fabry-Pérot mirror coatings***

The low loss mirror coatings were dielectric alternating-layer systems, each consisting of 28 non-quarter-wave layers. Given the specified spectral bandwidth, a non-zero dispersion was unavoidable. Zero dispersion was achieved by combining the two mirrors with precisely matched positive and negative dispersions, respectively. Ion beam sputtering was used for the precise fabrication of the coatings. The dispersion was subsequently measured using a white-light interferometer.

**References**

1. Borucki, W. J. et al. Kepler Planet-Detection Mission: Introduction and First Results. Science **327**, 977–980 (2010).
2. Ricker, G. R. et al. Transiting Exoplanet Survey Satellite. J. Astron. Telesc. Instrum. Syst **1**, 014003 (2014).
3. Rauer, H. et al. The PLATO mission. Exp Astron **59**:26 (2025).
4. Crass, J. et al. 2021, Extreme Precision Radial Velocity Working Group Final Report. arXiv:2107.14291.
5. Murphy, M. T. et al. High-precision wavelength calibration of astronomical spectrographs with laser frequency combs. Monthly Notices of the Royal Astronomical Society **380**, 839–847 (2007).
6. Quinlan, F., Ycas, G., Osterman, S. & Diddams, S. A. A 12.5 GHz-spaced optical frequency comb spanning >400 nm for near-infrared astronomical spectrograph calibration. Review of Scientific Instruments **81**, 063105 (2010).
7. Ycas, G. G. et al. Demonstration of on-sky calibration of astronomical spectra using a 25 GHz near-IR laser frequency comb. Opt. Express **20**, 6631 (2012).
8. Wilken, T. et al. A spectrograph for exoplanet observations calibrated at the centimetre-per-second level. Nature **485**, 611–614 (2012).
9. Li, C.-H. et al. A laser frequency comb that enables radial velocity measurements with a precision of 1 cm $s^{-1}$. Nature **452**, 610–612 (2008).
10. Braje, D. A., Kirchner, M. S., Osterman, S., Fortier, T. & Diddams, S. A. Astronomical spectrograph calibration with broad-spectrum frequency combs. Eur. Phys. J. D **48**, 57–66 (2008).
11. Glenday, A. G. et al. Operation of a broadband visible-wavelength astro-comb with a high-resolution astrophysical spectrograph. Optica **2**, 250 (2015).
12. McCracken, R. A. et al. Wavelength calibration of a high resolution spectrograph with a partially stabilized 15-GHz astrocomb from 550 to 890 nm. Opt. Express **25**, 6450 (2017).

13. Chae, E. et al. Compact green Ti:sapphire astro-comb with a 43 GHz repetition frequency. J. Opt. Soc. Am. B **38**, A1 (2021).
14. Yi, X. et al. Demonstration of a near-IR line-referenced electro-optical laser frequency comb for precision radial velocity measurements in astronomy. Nat Commun **7**, 10436 (2016).
15. Obrzud, E. et al. Broadband near-infrared astronomical spectrometer calibration and on-sky validation with an electro-optic laser frequency comb. Opt. Express **26**, 34830 (2018).
16. Metcalf, A. J. et al. Stellar spectroscopy in the near-infrared with a laser frequency comb. Optica **6**, 233 (2019).
17. Sekhar, P. et al. Tunable 30 GHz laser frequency comb for astronomical spectrograph characterization and calibration. Opt. Lett. **49**, 6257 (2024).
18. Suh, M.-G. et al. Searching for exoplanets using a microresonator astrocomb. Nature Photon **13**, 25–30 (2019).
19. Obrzud, E. et al. A microphotonic astrocomb. Nature Photon **13**, 31–35 (2019).
20. Ludwig, M. et al. Ultraviolet astronomical spectrograph calibration with laser frequency combs from nanophotonic lithium niobate waveguides. Nat Commun **15**, 7614 (2024).
21. Paschotta, R. Noise of mode-locked lasers (Part II): timing jitter and other fluctuations. Appl Phys B **79**, 163–173 (2004).
22. Benedick, A. J. et al. Visible wavelength astro-comb. Opt. Express **18**, 19175 (2010).
23. Phillips, D. F. et al. Calibration of an astrophysical spectrograph below 1 m/s using a laser frequency comb. Opt. Express **20**, 13711 (2012).
24. Cheng, Y. S. et al. Continuous ultraviolet to blue-green astrocomb. Nat Commun **15**, 1466 (2024).
25. Roth, P. W., Maclean, A. J., Burns, D. & Kemp, A. J. Direct diode-laser pumping of a mode-locked Ti:sapphire laser. Opt. Lett. 36, 304 (2011).
26. Durfee, C. G. et al. Direct diode-pumped Kerr-lens mode-locked Ti:sapphire laser. Opt. Express 20, 13677 (2012).
27. Castro-Marin, P., Mitchell, T., Sun, J. & Reid, D. T. Characterization of a carrier-envelope-offset-stabilized blue- and green-diode-pumped Ti:sapphire frequency comb. Opt. Lett. 44, 5270 (2019).

28. Ostapenko, H., Mitchell, T., Castro-Marin, P. & Reid, D. T. Three-element, self-starting Kerr-lens-modelocked 1-GHz Ti:sapphire oscillator pumped by a single laser diode. Opt. Express **30**, 39624 (2022).
29. Ostapenko, H., Mitchell, T., Castro-Marin, P. & Reid, D. T. Design, construction and characterisation of a diode-pumped, three-element, 1-GHz Kerr-lens-modelocked Ti:sapphire oscillator. Appl. Phys. B **129**, 33 (2023).
30. Allan, E. et al. Octave-spanning frequency comb from a single-diode-pumped 1 GHz Ti:sapphire laser. Opt. Lett. **51**, 337 (2026).
31. Allan, E. et al. Towards a single-laser-diode-pumped 15-GHz Ti:sapphire astrocomb. Opt. Express **34**, 6025 (2026).
32. Bartels, A. et al. 10-GHz self-referenced optical frequency comb. Science **326**, 681 (2009).
33. Bauch, A. Caesium atomic clocks: function, performance and applications. Meas. Sci. Technol. **14**, 1159–1173 (2003).
34. Telle, H. R. et al. Carrier-envelope offset phase control: A novel concept for absolute optical frequency measurement and ultrashort pulse generation. Appl Phys B **69**, 327–332 (1999).
35. Alabbadi, A. et al. Visible octave frequency combs in silicon nitride nanophotonic waveguides driven by Ti:sapphire lasers. arXiv:2601.04047 (2026).
36. Barnes, S. I. et al. The optical design of the Southern African large telescope high resolution spectrograph: SALT HRS. SPIE Proceedings **7014**, 70140K (2008).
37. Bramall, D. G. et al. The SALT HRS spectrograph: instrument integration and laboratory test results. SPIE Proceedings **8446**, 84460A (2012).
38. Cheng, Y. S. et al. Comb-mode sweeping in a 650 nm–1030 nm astrocomb. Opt. Express **34**, 8569 (2026).
39. Kirchner, M. S. A low-threshold self-referenced Ti:Sapphire optical frequency comb. Opt. Express **14**, 9531-9536 (2006).

**Funding sources**

This research was supported by UK Research and Innovation and the Royal Academy of Engineering (grant numbers EP/Y011422/1, ST/V000403/1, ST/Y001273/1, ST/X004503/1 and RCSRF2223-1678 to D.T.R.); the Deutsche Forschungsgemeinschaft (grant number 541267874 to P.D.), the Max Planck School of Photonics, and the German Federal Ministry of Research, Technology and Space (grant numbers 13N17314, 13N17342 to P.D.).

**Author contributions**

D.T.R. conceived the study and wrote the original manuscript draft. H.O., P.C.M. and E.A. designed and built the laser. A.A. designed and fabricated the silicon nitride waveguides and P.D. provided the waveguides. E.A., D.T.R., M.C.S., D.L.H. and L.A.C. performed the experiments at SALT and carried out data analysis. Y.S.C. designed the Fabry-Pérot cavity, stabilisation and beat-detection unit. R.A.M. contributed to the specification of the Fabry-Pérot mirrors and delivery fibre. K.D., W.N. and J.M.C. contributed to the astrocomb preparation and characterisation. T.G. and T.W. designed the dispersion-engineered Fabry-Pérot mirror coatings, which were manufactured by Laseroptik. All authors reviewed and approved the final manuscript.

**Competing interests statement**

The authors declare no competing interests.

**Figures**

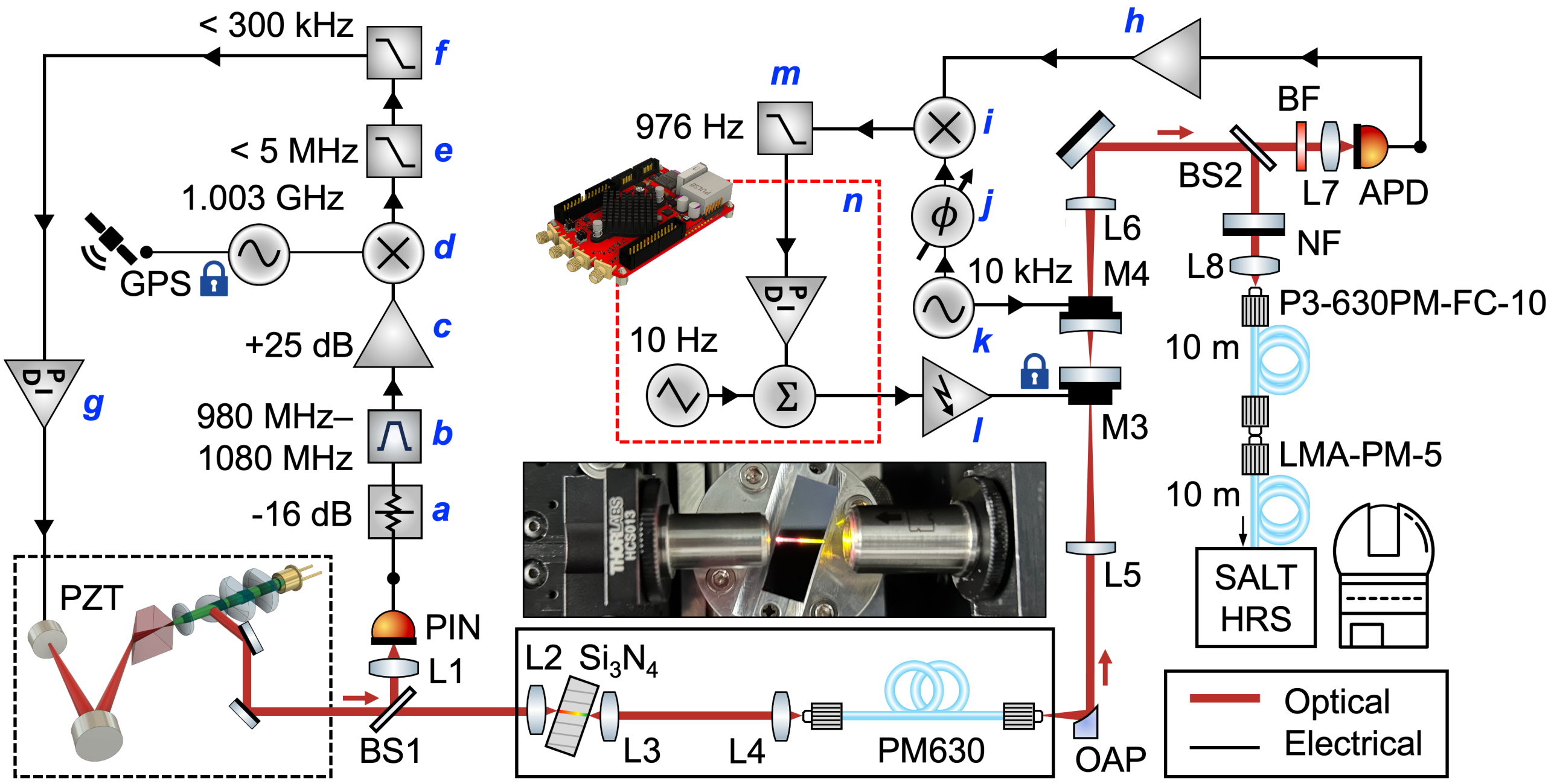

**Fig. 1**. Astrocomb configuration. The 1 GHz Kerr-lens-modelocked diode-pumped Ti:sapphire laser (dashed box) pumps a $Si_3N_4$ waveguide to create a supercontinuum that is fibre coupled into a 21 GHz Fabry-Pérot filter cavity before fibre delivery to the SALT HRS spectrograph. **a**–**g**, Comb-mode-spacing stabilisation servo loop. **h**–**n**, Fabry-Pérot length-stabilisation servo loop. See main text for other details.

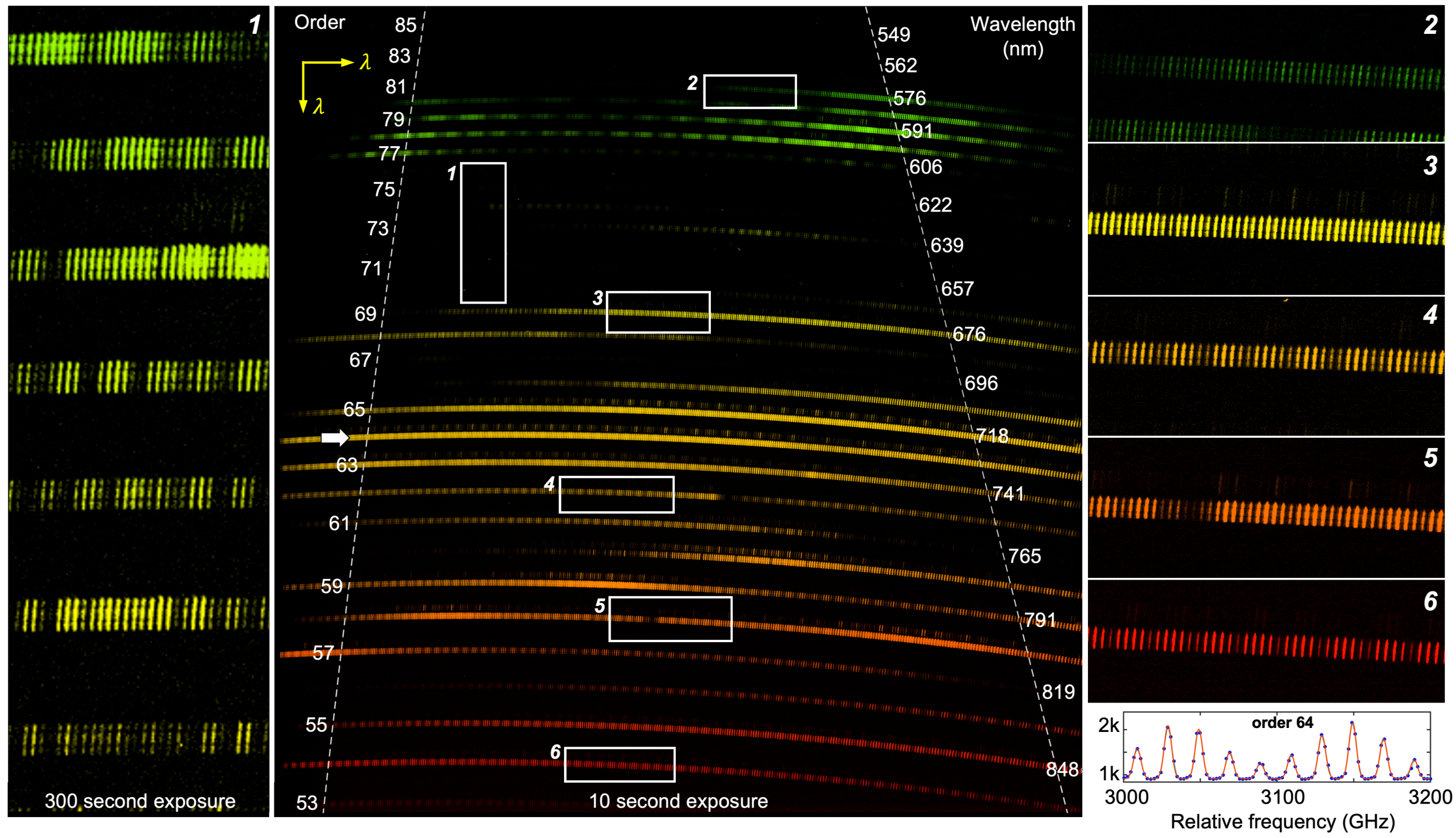

**Fig. 2**. Image of the 21 GHz astrocomb observed on the 550 nm to 890 nm red channel of SALT-HRS, showing resolved comb modes across most of the available orders. The main frame is acquired with a 10 second exposure time, while the insets show the details of individual regions at exposure times of 300 seconds (1) and 10 seconds (2–6). A line-cut of order 64 (light blue arrow) shows an example of the Gaussian fits to ten comb modes.

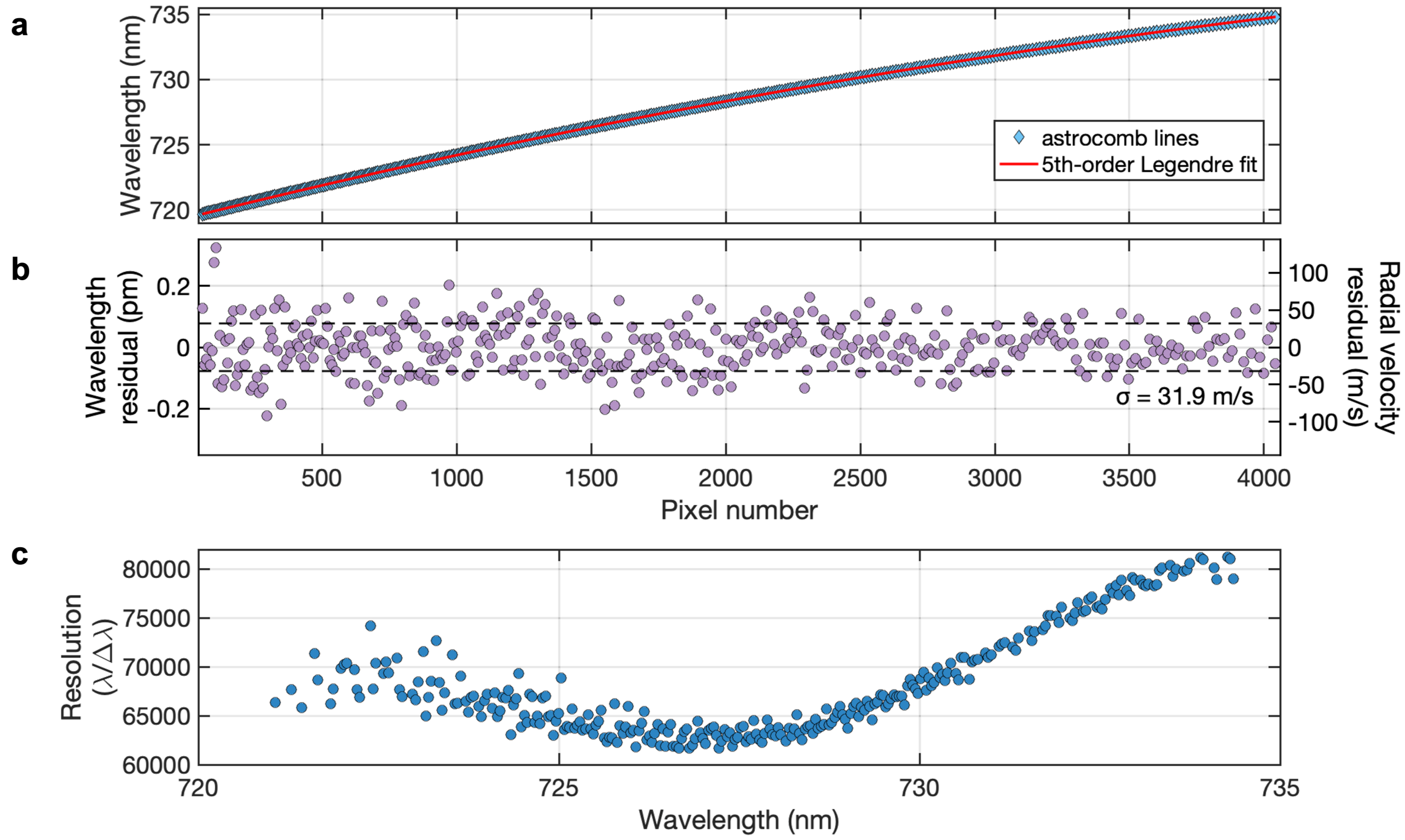

**Fig. 3**. Analysis of the astrocomb detected on order 64 of the SALT HRS spectrograph. **a**, Wavelength assignation of the astrocomb modes observed across the detector (diamonds) and corresponding wavelength solution (red line) from a fifth-order Legendre polynomial fit. **b**, Fitting residuals expressed in wavelength (left axis) and radial-velocity (right axis), indicating a radial-velocity precision of 31.9 m $s^{-1}$. **c**, Wavelength-dependent resolving power inferred from Gaussian line-spread-function fits to each of the observed comb-modes.

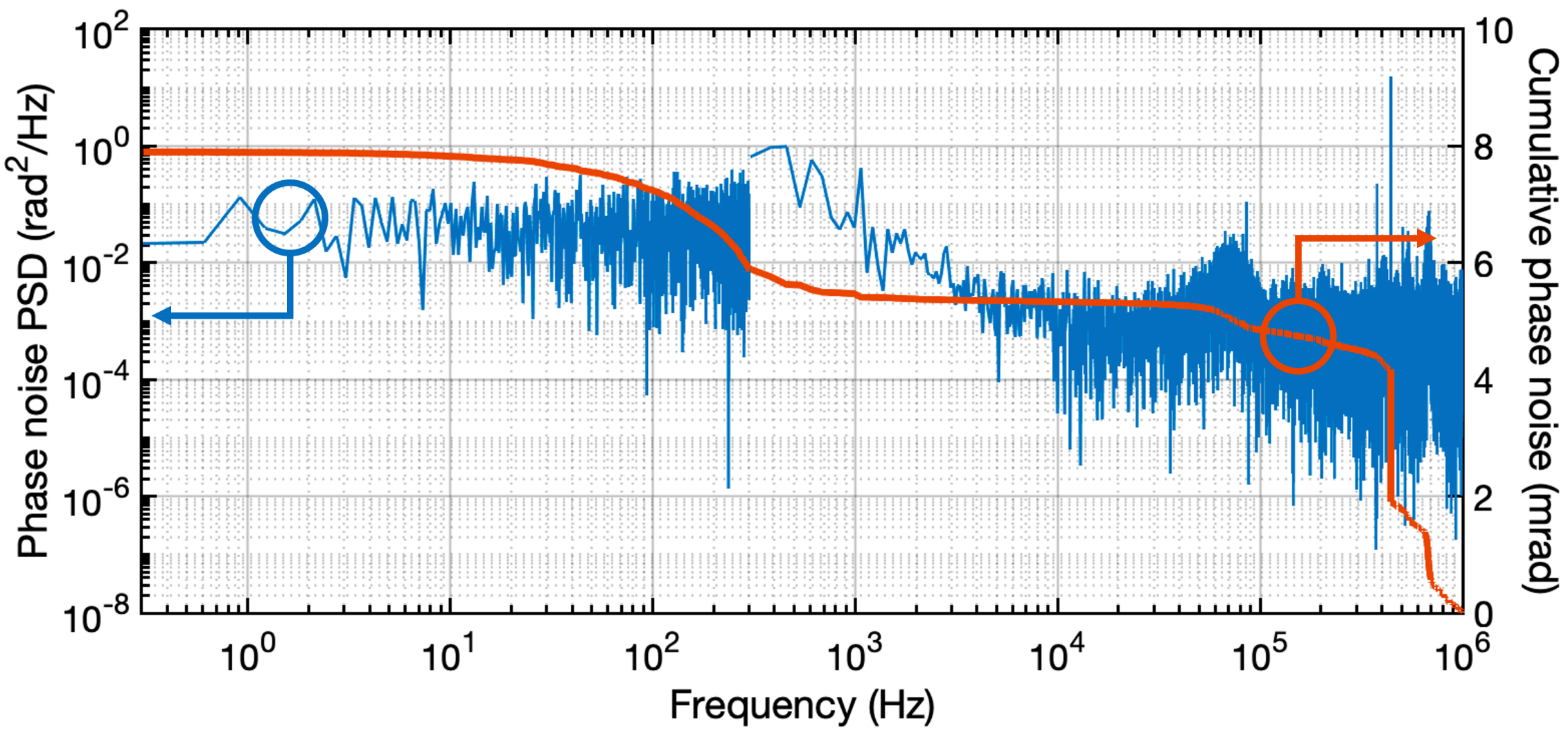

**Fig. 4**. Stability of the comb-mode spacing. In-loop phase-noise power spectral density of the $f_{\mathrm{rep}}$ servo-loop error signal when $f_{\mathrm{rep}}$ was stabilized to a GPS-disciplined reference oscillator at 1,003,004,600 Hz. The data are reconstructed from a 3.3-second measurement recorded at 20 kHz sampling rate and a 13-ms measurement recorded at 5 MHz sample rate. The phase noise integrated from 1 MHz to 300 mHz (right axis) is 8 mrad.

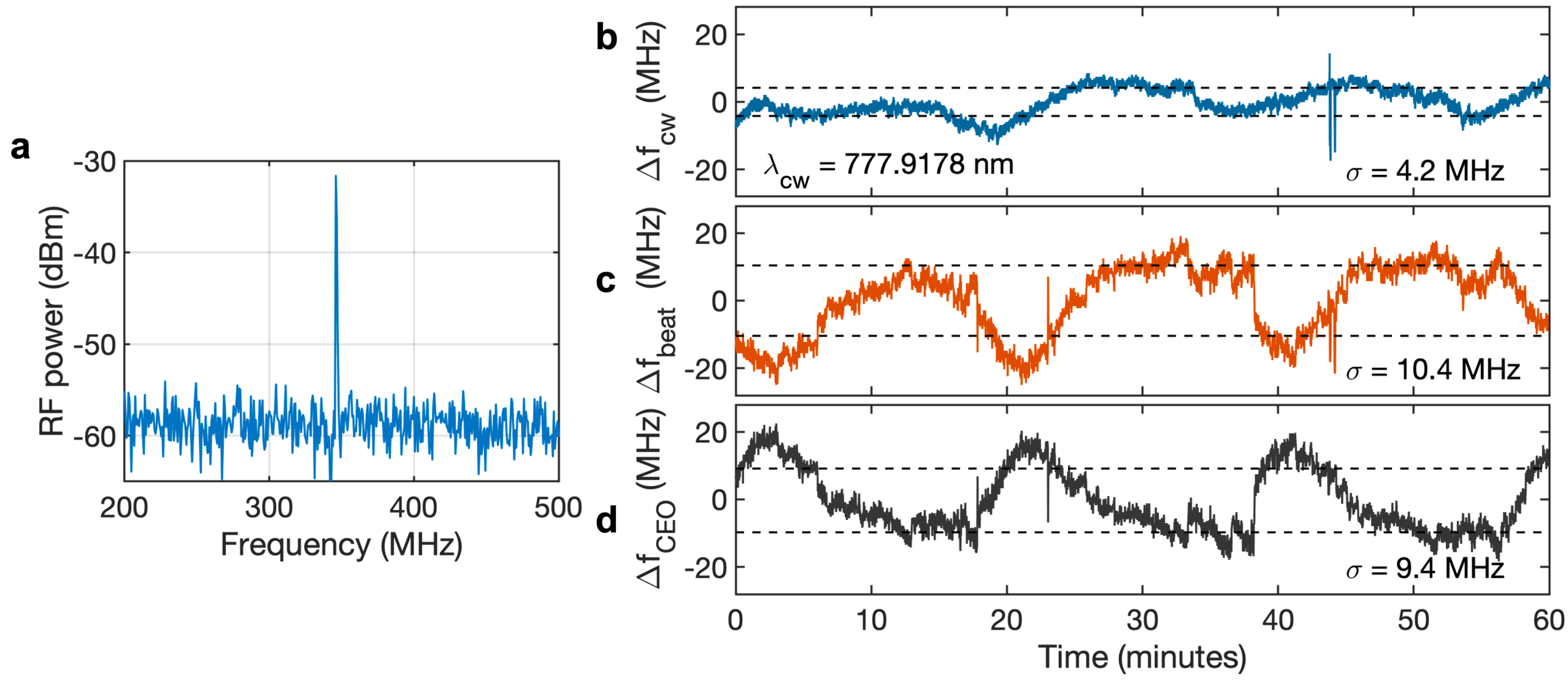

**Fig. 5**. Evaluation of the comb offset stability over one hour when $f_{\mathrm{rep}}$ is locked. **a**, Example heterodyne beat frequency between the comb and a single-frequency laser at 777.9178 nm. **b**, Frequency variations of a 780 nm single-frequency laser and **c**, simultaneously recorded changes in the heterodyne beat frequency ($f_{\mathrm{cw}} - f_n$) between the comb and the cw laser. **d**, Difference between the beat and cw laser frequencies, corresponding to fluctuations in the comb offset frequency, $f_{\mathrm{CEO}}$. In **b** and **c** the mean has been subtracted from the data before plotting. Annotation shows the rms variation ($\sigma$).

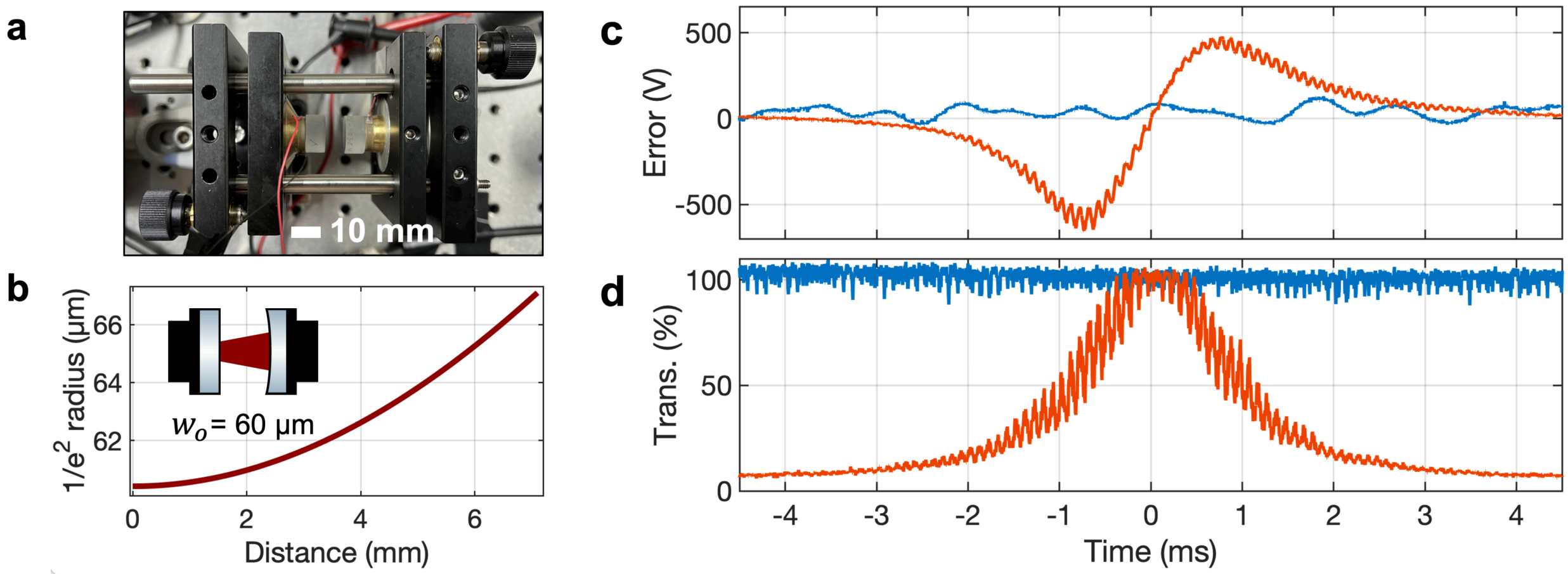

**Fig. 6**. Fabry-Pérot filter cavity design and performance. **a**, Mechanical and **b**, optical design, showing the mirror separation of 7.12 mm and a waist of 60 μm $1/e^2$ radius formed on the plane mirror. **c**, In-loop locked (blue) and unlocked (orange) servo-loop error signal, and **d**, corresponding transmission of the supercontinuum comb with a mode spacing of $21f_{\text{rep}}$ (21.0631 GHz).

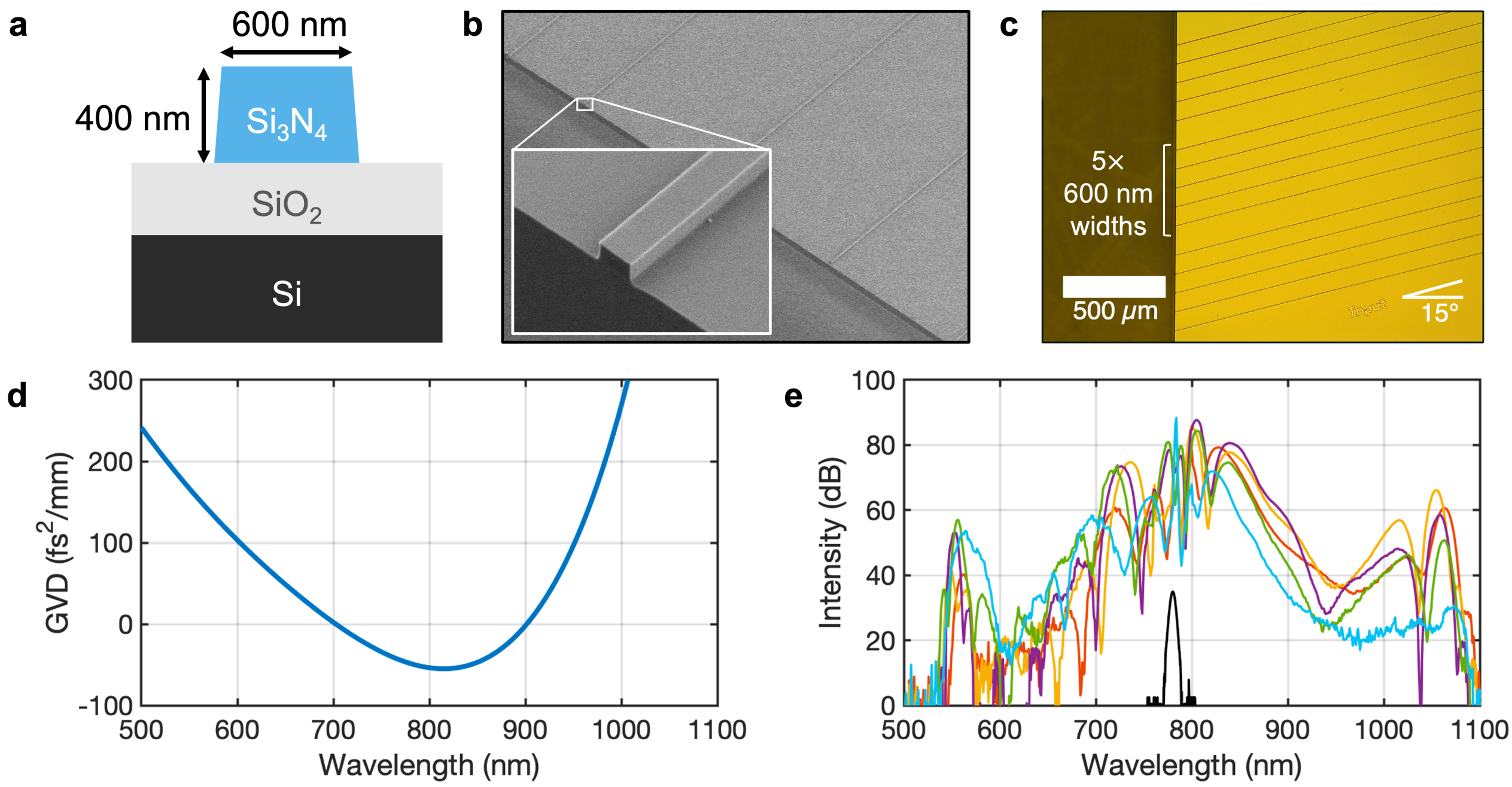

**Fig. 7**. Silicon nitride waveguide design and performance. **a**, Schematic of the cross section and dimensions of the air-clad $Si_3N_4$ waveguide. **b**, Scanning electron microscope image of the chip with silicon nitride waveguides and a zoom into the coupling facet at the side of the chip (inset). **c**, Optical microscope image, showing the 15° facet angle of the tilted waveguides. **d**, Dispersion profile of the quasi-TE mode. **e**, Supercontinuum spectra from each waveguide in the 600 nm-width group, pumped by 170 mW at 1 GHz (estimated 25 pJ on-chip energy).

**SUPPLEMENTARY INFORMATION**

# Demonstration of a Single-Laser-Diode-Pumped Ti:Sapphire Astrocomb on the Southern African Large Telescope

Ewan Allan[1], Yuk Shan Cheng[1], Kamalesh Dadi[1], Pablo Castro-Marin[1], Jake M. Charsley[1], William Newman[1], Abdullah Alabbadi[2,3], Hanna Ostapenko[1], Richard A. McCracken[1], Pascal Del'Haye[2,4], Thomas Willemsen[5], Tobias Gross[5], Daniel L. Holdsworth[6,7,8], Malcolm C. Scarrott[6,7], Lisa A. Crause[6,7], and Derryck T. Reid[1]*

*D.T.Reid@hw.ac.uk

1. Institute of Photonics and Quantum Sciences, Heriot-Watt University, Edinburgh, UK.
2. Max Planck Institute for the Science of Light, Erlangen, Germany.
3. Department of Physics, Alexandria University, Alexandria, Egypt.
4. Department of Physics, Friedrich-Alexander-Universität Erlangen-Nürnberg, Germany.
5. LASEROPTIK GmbH, Johannes-Ebert-Str. 1, 30826 Garbsen, Germany.
6. South African Astronomical Observatory, Cape Town, South Africa.
7. Southern African Large Telescope, South Africa.
8. School of Physics, Engineering and Technology, University of York, Heslington, York, UK.

**Supplementary Note 1:**

**$Si_3N_4$ waveguide design and visible dispersive-wave generation**

The $Si_3N_4$ waveguide used for supercontinuum generation consisted of an 8-mm-long air-clad nonlinear section with a fixed height of 400 nm and a uniform width of 600 nm. This geometry was chosen to provide anomalous dispersion around the Ti:sapphire pump wavelength as shown in Fig. S1b, enabling soliton-mediated spectral broadening and dispersive-wave generation toward both the visible and near-infrared [1]. To improve input and output coupling, the waveguide width was tapered to 2.5 µm near the chip facets as demonstrated in Figure S1a.

We used the generalized nonlinear Schrödinger equation [2,3] to evaluate how the input pulse duration affects the nonlinear broadening dynamics under experimentally relevant waveguide conditions and including measured propagation losses of 3.6 dB/cm [1]. In Figure S1c, the input peak power was kept fixed at approximately 290 W while the pulse duration was varied. Coherent octave-spanning supercontinuum is feasible for pulse durations shorter than 140 fs (see Supplementary Note 2). The visible dispersive-wave integrated power is shown in Fig.S1d, and is maximized for pulse durations in the few-tens-of-femtoseconds range and remains high near the experimentally used value of 60 fs. Driving the anomalous dispersion $Si_3N_4$ waveguide close to the visible band ensures higher conversion efficiency into the visible dispersive-wave than the near-infrared dispersive wave as shown in Fig. S1d. For longer pulses, particularly above ~100 fs, the spectral energy transferred to both dispersive-wave bands decreases strongly due to weaker spectral overlap of the higher-order soliton and phase-matched dispersive-wave wavelength.

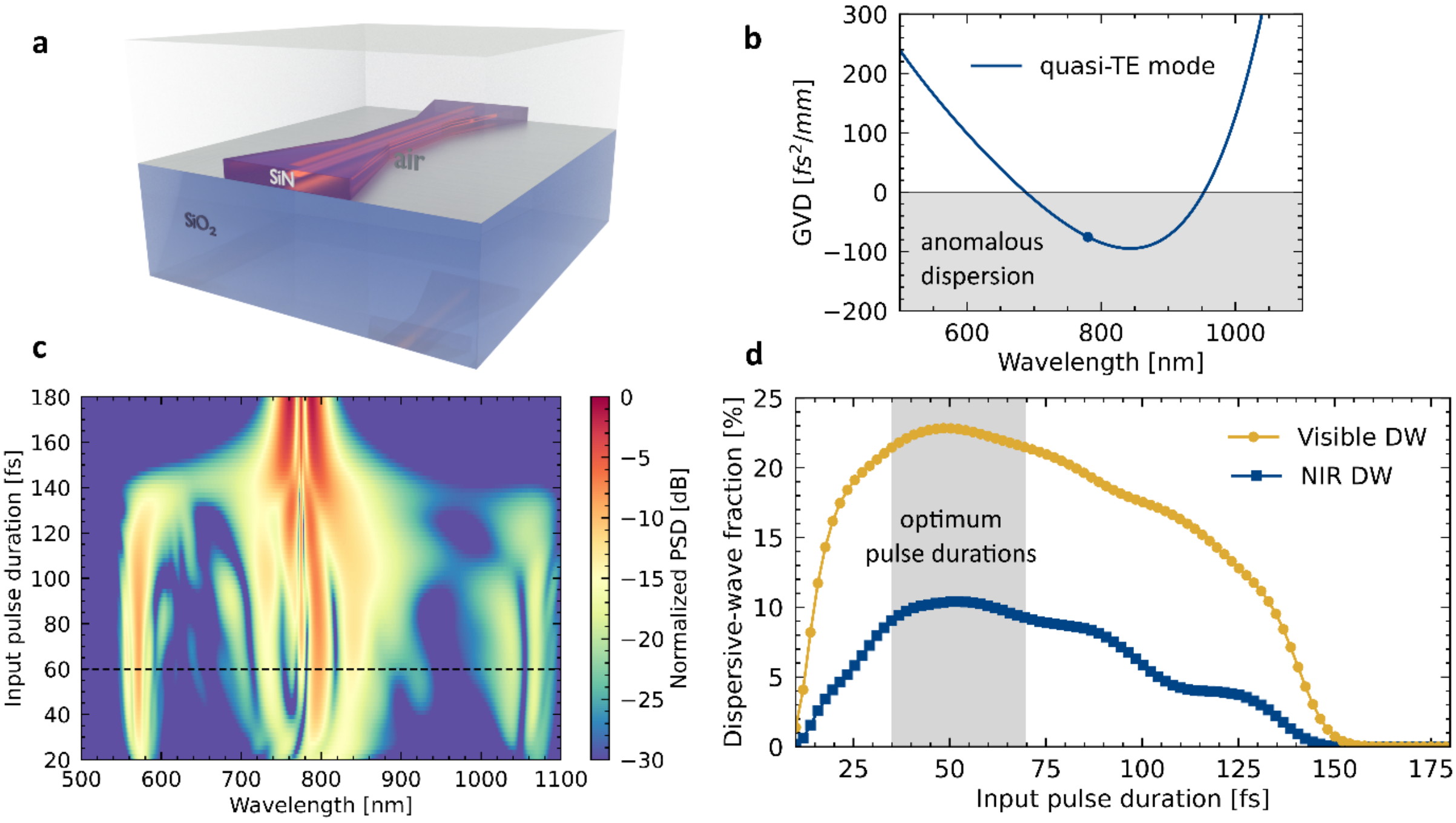

**Fig. S1. Pulse-duration dependence of dispersive-wave generation in the $Si_3N_4$ waveguide. a**, Schematic of the air-clad $Si_3N_4$ waveguide used for visible supercontinuum generation. **b**, Calculated group-velocity dispersion of the quasi-TE mode for 600 nm × 400 nm waveguide, showing anomalous dispersion across the pump region and suitable for dispersive-wave generation toward the visible and near-infrared. **c**, Simulated output power spectral density as a function of input pulse duration, calculated using experimentally relevant peak power of ~300 W and waveguide parameters. The dashed line represents the condition used in this work. **d**, Integrated spectral-energy fraction transferred to the visible and near-infrared dispersive-wave bands as a function of input pulse duration. The shaded region marks the short-pulse operating range in which visible dispersive-wave generation is most efficient. The experimentally used pulse duration of 60 fs lies close to the optimum for visible conversion.

**Supplementary Note 2:**

**Coherent operating regime for visible-to-near-infrared supercontinuum generation**

The pulse duration used in our experiment is optimum for maximizing the power transfer to visible band. However, it is crucial that the generated octave supercontinuum maintains its coherence. The output coherence of the supercontinuum was evaluated using an ensemble of independent stochastic nonlinear propagation simulations. For each input pulse duration, M = 200 simulations were performed with identical deterministic parameters and independent quantum-vacuum noise seeds added to the input field. The wavelength-dependent first-order spectral coherence was calculated from the ensemble following the standard approach used in noise-seeded supercontinuum [4,5]. The ensemble-mean output spectrum is shown in Fig. S2b and was calculated as

$$S_{\text{mean}}(\lambda) = \frac{1}{M}\sum_{i=1}^{M} |A_i(c/\lambda)|^2 \frac{c}{\lambda^2}$$

The first-order spectral coherence at zero path delay is defined as

$$g_{12}^{(1)}(\lambda) = \frac{\langle A_1^*(\lambda) A_2(\lambda)\rangle}{\sqrt{\langle |A_1(\lambda)|^2\rangle\langle |A_2(\lambda)|^2\rangle}}.$$

The coherence result shown in Fig. S2c is the modulus $\left|g_{12}^{(1)}(\lambda)\right|$, which ranges from 0 to 1, with 1 corresponding to identical shot-to-shot spectral amplitude and phase. To quantify the coherence in selected spectral regions, we calculated a band-averaged spectral coherence over a wavelength window $W = [\lambda_1, \lambda_2]$. The averaged coherence was defined as

$$\left\langle \left|g_{12}^{(1)}\right| \right\rangle_W = \frac{\int_{\lambda_1}^{\lambda_2} \left|g_{12}^{(1)}(\lambda)\right| S_{\text{mean}}(\lambda) d\lambda}{\int_{\lambda_1}^{\lambda_2} S_{\text{mean}}(\lambda) d\lambda},$$

where $S_{mean}(\lambda)$ is the ensemble-averaged output spectral density. This weighting gives the coherence of the spectral components that carry optical energy in the selected band [5]. Fig. S2a confirms that the short-pulse regime used in the experiment preserves high coherence across the spectral bands relevant for comb generation. The wavelength-resolved coherence results in Figs. S2b and S2c confirm that coherence is maintained across most of the generated supercontinuum at this operating point, with reduced values appearing mainly for longer input pulses and in weak spectral regions near the edges of the supercontinuum. In particular, the 60-fs operating point lies within a broad high-coherence window, where the pump region, visible dispersive wave, and near-infrared dispersive wave all show near-unity coherence.

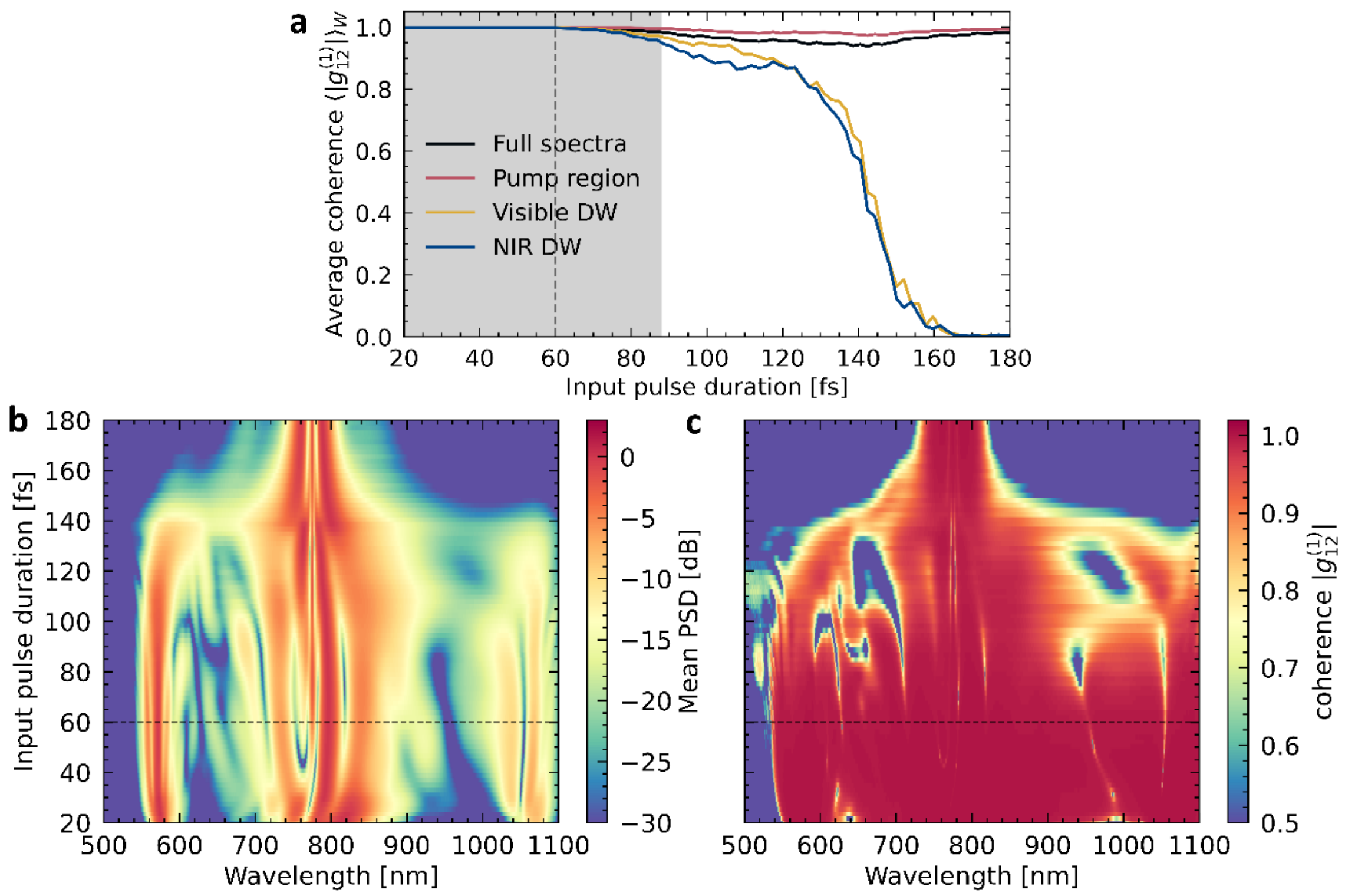

**Fig. S2. Pulse-duration operation window for coherent $Si_3N_4$ supercontinuum generation. a**, Spectrally weighted average coherence $\left\langle \left| g_{12}^{(1)} \right| \right\rangle_w$, calculated as a function of input pulse duration for the full simulated spectrum, the pump region, the visible dispersive-wave band, and the near-infrared dispersive-wave band. The spectral windows used for the averaging are 540 nm to 650 nm for the visible dispersive wave, 720 nm to 850 nm for the pump region, and 960 nm to 1100 nm for the near-infrared dispersive wave. The shaded area marks the short-pulse regime in which the dispersive-wave bands maintain high coherence >0.95, and the dashed vertical line indicates the 60-fs pulse duration used in the experiment. **b**, Mean output power spectral density from 200 independent noise-seeded pulses for different input pulse duration. **c,** Corresponding wavelength-resolved coherence $\left| g_{12}^{(1)} \right|$ showing that the 60-fs operating point lies within a high-coherence regime across the pump and dispersive-wave spectral bands.

**Supplementary Note 3:**

**Simulation-experiment comparison of $Si_3N_4$ octave generation**

We compared the nonlinear propagation model with experimental observations of the $Si_3N_4$ waveguide output. The simulated spectral evolution in Fig. S3a shows that the launched Ti:sapphire pulse first undergoes soliton-mediated broadening around the pump wavelength. After approximately 4-4.5 mm of propagation, spectral energy is transferred to dispersive waves on both sides of the pump, producing visible and near-infrared spectral components. The camera image of the waveguide during operation in Fig. S3b provides a qualitative spatial signature of this process. The near-infrared pump light is not visible to the camera in the initial section of the waveguide, whereas visible scattered light appears farther along the propagation direction, at a location consistent with the simulated onset of visible dispersive-wave generation. The dispersive-wave positions were estimated using the resonant-radiation phase-matching condition. In the soliton co-moving frame, the phase mismatch is

$$\Delta\beta(\omega) = \beta(\omega) - \beta(\omega_s) - \beta_1(\omega_s)(\omega - \omega_s) - \Delta\beta_{\mathrm{NL}},$$

where $\omega_s$ is the soliton centre frequency, $\beta_1(\omega_s)$ is the inverse group velocity of the soliton, and $\Delta\beta_{\mathrm{NL}}$ is the nonlinear contribution to the soliton propagation constant. The dispersive-wave wavelengths are obtained from the zero-crossings of $\Delta\beta(\omega)$. The calculated phase-matching curve in Fig. S3c predicts visible and near-infrared dispersive-wave bands in the wavelength regions where spectral peaks are observed experimentally. Fig. S3d shows the measured output spectra from five nominally identical waveguides on the same chip show reproducible visible-to-near-infrared supercontinuum generation under the same pumping conditions. The main spectral features agree with the phase-matching calculation and with the simulated spectral evolution, supporting the assignment of the visible and near-infrared bands to dispersive-wave emission. At the experimental operating point, the deterministic simulation and the ensemble-mean spectrum from noise-seeded simulations are nearly identical. The calculated first-order spectral coherence remains close to unity across the spectral regions used for astrocomb operation, with coherence dips appearing mainly in low-power spectral regions where the spectral density is below approximately -30 dB.

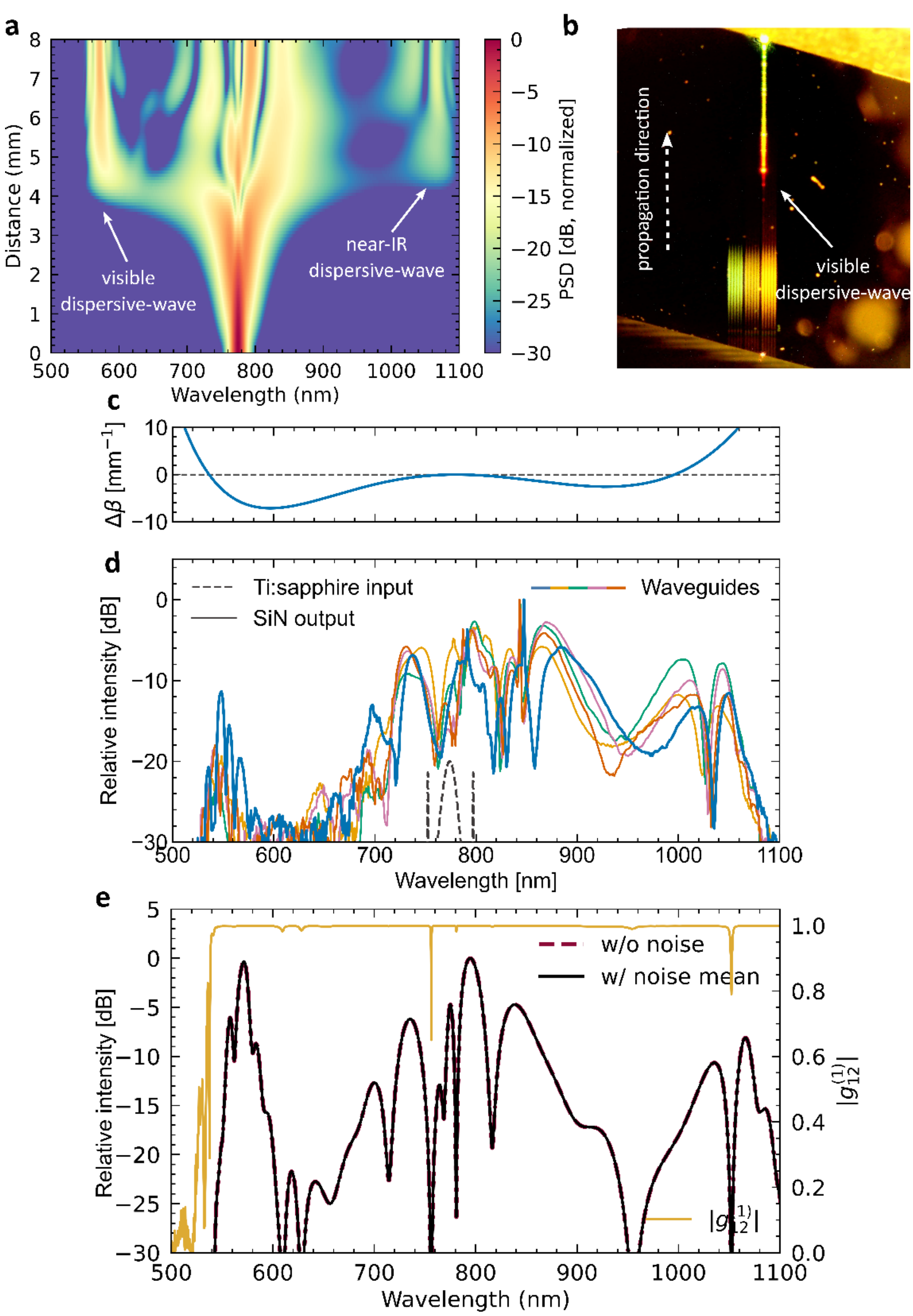

**Fig. S3. Simulation-experiment comparison of dispersive-wave generation in the $Si_3N_4$ waveguide. a**, Simulated spectral evolution along the 8-mm-long $Si_3N_4$ waveguide for the experimentally used pulse parameters. The calculation shows soliton-mediated broadening followed by the emergence of visible and near-infrared dispersive waves after approximately 4–4.5 mm of propagation. **b**, Camera image of the waveguide during operation, with the propagation direction indicated by the dashed arrow. The near-infrared pump is not detected by the camera in the initial section of the waveguide, while visible scattered light appears farther along the propagation direction. This onset is qualitatively consistent with the simulated build-up of the visible dispersive-wave in a. **c**, Calculated dispersive-wave phase mismatch $\Delta\beta$ for the quasi-TE mode of the waveguide. The zero-crossings indicate the estimated phase-matched wavelengths for dispersive-wave emission on the visible and near-infrared sides of the Ti:sapphire pump. **d**, Measured output spectra from five nominally identical waveguides on the same chip under the same pumping conditions. The input Ti:sapphire spectrum is shown as a dashed line and shifted vertically by 20 dB for clarity. The measured visible and near-infrared spectral features are consistent with the calculated phase-matching condition. **e**, Simulated output spectrum and first-order spectral coherence at the experimental operating point. The dashed line shows the deterministic simulation without input noise, while the solid black line shows the ensemble-mean spectrum from 200 independent noise-seeded simulations. The corresponding $\left|g_{12}^{(1)}\right|$ remains close to unity across the spectral regions relevant to the astrocomb, with reduced values occurring mainly where the spectral density is below approximately −30 dB.

## Supplementary references

1. Alabbadi, A. et al. Visible octave frequency combs in silicon nitride nanophotonic waveguides driven by Ti:sapphire lasers. arXiv:2601.04047 (2026).
2. J. M. Dudley, G. Genty and S. Coen, "Supercontinuum generation in photonic crystal fiber," Reviews of Modern Physics **78**, 1135–1184 (2006).
3. J. Hult, "A fourth-order Runge–Kutta in the interaction picture method for simulating supercontinuum generation in optical fibers," Journal of Lightwave Technology **25**, 3770 (2007).
4. J. M. Dudley and S. Coen, "Coherence properties of supercontinuum spectra generated in photonic crystal and tapered optical fibers," Optics Letters **27**, 1180–1182 (2002).
5. A. M. Heidt, J. S. Feehan and J. H. V. Price, "Limits of coherent supercontinuum generation in normal dispersion fibers," JOSA B **34**, 764–775 (2017).